# Contactless Modulation of Intralayer and Interlayer Excitons in MoS$_2$/WSe$_2$ heterostructures with Acoustoelectric Fields

*Yueyi Sun[†], Dexing Liu[†], Jiefei Zhu, Siming Liu, Jiwei Chen, Yingjie Luo, Yihong Sun, Mansun Chan, Cary. Y. Yang, Taojie Zhou, Min Zhang\* and Changjian Zhou\**

Yueyi Sun, S. Liu, J. Chen, Y. Luo, Yihong Sun, T. Zhou, C. Zhou

School of Microelectronics, South China University of Technology, Guangzhou 511442, P. R. China

D. Liu, M. Zhang

School of Science and Engineering, The Chinese University of Hong Kong, Shenzhen, Shenzhen 518172, P. R. China

D. Liu, J. Zhu

School of Electronic and Computer Engineering, Peking University, Shenzhen 518055, P. R. China

M. Chan

Department of Electronic and Computer Engineering, The Hong Kong University of Science and Technology, Kowloon, Hong Kong, P. R. China

C. Y. Yang

Center for Nanostructures, Santa Clara University, Santa Clara, California, USA

\*Corresponding authors Email: zhoucj@scut.edu.cn (Changjian Zhou) ; mzhang@cuhk.edu.cn (Min Zhang)

[†]Yueyi Sun and Dexing Liu contributed equally to this work.

**Abstract**

This work presents a platform that enables surface acoustic wave (SAW) modulation of both intralayer and interlayer excitons in $MoS_2/WSe_2$ heterostructures. Harnessing the coupled piezoelectric and strain fields of SAWs, this integrated approach allows for dynamic, precise, and fully contactless control of excitonic properties, a capability essential for the realization of next-generation optoelectronic, quantum photonic, and excitonic devices. We identify two distinct modulable interlayer excitons in optical communication bands: $IX_{K-\Gamma}$ in the O-band (around 1300 nm) and $IX_{K-K}$ in the S-band (around 1500 nm), these two excitons display a robust twist-angle-independent energy splitting of 120 meV, in agreement with density functional theory (DFT) calculations. The type-II band alignment induced by the SAW not only promotes efficient exciton dissociation but also enables direct and tunable modulation of photoluminescence via the formation of confined piezoelectric potential wells. Furthermore, by simultaneously generating in-plane and out-of-plane SAW fields, the platform achieves selective manipulation of intralayer and interlayer excitons, inducing quadratic Stark effects for intralayer excitons and linear Stark effects for interlayer excitons. These findings provide new insights into SAW–exciton interactions in van der Waals heterostructures, broaden the operational spectral range, and establish pathways toward on-chip, acousto-optic, and quantum optoelectronic devices with advanced excitonic functionality.

## 1. Introduction

Transition metal dichalcogenide (TMD) monolayers and their stacked heterostructures have emerged as promising material platforms for future optoelectronic applications owing to their moderate bandgaps, outstanding optical properties, as well as multi-exciton and spin-valley characteristics[1-3]. The optical properties of these materials are primarily dominated by excitons, which exhibit large oscillator strengths and robust interband charge excitations[4]. To date, a variety of methods, including magnetic fields[5,6], optical fields[7], pressure[8,9], and electric fields[10], have been utilized to externally modulate the carrier and exciton optical emissions in these systems. Among these approaches, surface acoustic wave (SAW) enables non-contact driving *via* piezoelectric potential, allowing precise control of carriers and excitons transport in semiconductor nanostructures (e.g., quantum wells, nanowires, quantum dots)[11-13] and 2D materials[14-17]. However, existing research on SAWs modulation of TMD excitons has predominantly focused on intralayer excitons in monolayer materials[2],[14],[18-22]. Although interlayer excitons have been explored in some homostructure multilayer systems under SAW modulation[20],[22,23], the dynamic manipulation of interlayer excitons in heterostructures, which possess advantageous properties such as longer lifetimes, strong electric-field tunability, and controllable valley degrees of freedom, remains largely unexplored. Moreover, the simultaneous and selective control of both intralayer and interlayer excitons within heterostructures, and a comprehensive understanding of their distinct interactions with SAWs, is still lacking. More in-depth understanding is crucial for envisioning advanced applications such as integration with photonic circuits or quantum communication[24].

We present a platform that enables SAW modulation of both intralayer and interlayer excitons in $MoS_2$/$WSe_2$ heterostructures. We identify experimentally two distinct interlayer excitons in the near short-wave infrared (nSWIR) spectral region, i.e., $IX_{K\text{-}\Gamma}$ in the O-band (1300 nm), and $IX_{K\text{-}K}$ in the S-band (1500 nm). Although multiple photoluminescence (PL) features in the nSWIR range have been observed previously in $MoS_2$/$WSe_2$ heterobilayers, their valley origins have not been explicitly resolved[25-27]. In this work, we achieve a clear valley-resolved identification through combined experimental measurements and DFT calculations, attributing the ~120 meV splitting to transitions between distinct momentum valleys. This splitting remains robust over a wide range of twist angles, and the two excitons exhibit different thermal stability and responses to SAW modulation. We also observed indirect excitons $X_I^0$ and $X_I^-$ arising from transitions between the K and hybridized Γ valley within the visible (VIS) and near-infrared (NIR) range. The emission wavelengths of the observed interlayer excitons, which span the communication O-band and S-band, suggest their strong potential for seamless integration with on-chip optoelectronic devices for optical signal generation, transmission, and processing[28].

By exploiting the pronounced band alignment from type-II SAW modulation, we further promote exciton dissociation and generate electron-hole pairs confined in piezoelectric potential wells, enabling tunable modulation of exciton photoluminescence (PL) intensity. In particular, the in-plane and out-of-plane piezoelectric fields generated by SAWs allow simultaneous and contactless control of excitonic properties, leading to the observation of a quadratic Stark effect for intralayer excitons and a linear Stark effect for interlayer excitons. Unlike conventional electrostatic gating[29], this SAW-based approach is contactless, thus preventing material damage and interface contamination, and eliminating defects associated with electrode fabrication. We present a detailed comparison of our work's novelty with relevant studies in Supplementary Note 1. These results provide critical theoretical and experimental insights for the advancement of high-performance exciton devices based on TMD heterostructures.

## 2. Results and Discussion

### 2.1. Exciton Spectra and SAW Electric Field

**Figure 1**a illustrates the schematic diagram of the proposed platform. The SAWs generated by the interdigital electrodes propagate along the x-axis (as indicated by the arrow). We transfer monolayers of $WSe_2$ and $MoS_2$ grown by chemical vapor deposition onto a 128° Y-X cut $LiNbO_3$ substrate with a twist angle of approximately 3° (confirmed by second-harmonic generation (SHG) measurements, Supplementary Figure 1a), as shown in Figure 1b. $WSe_2$ is in direct contact with the substrate, and $MoS_2$ is stacked on top of it with an AB stacking (Supplementary Note 2 and Supplementary Figure 1b). The transfer position is precisely aligned with the propagation path of the SAWs, enabling efficient modulation of excitonic properties by the acoustic waves, the relative position is shown in Supplementary Figure 2 and Supplementary Note 3. Optical excitation is carried out using a 532 nm laser beam focused to a spot size of 2.6 μm, we use an excitation power of 100 μW, which can avoid high exciton density effects such as screening or nonlinear SAW-assisted dissociation (Supplementary Note 4 and Supplementary Figure 3). By varying the excitation intensity of the SAWs, we monitor changes in the PL emission spectra. By increasing radio frequency (RF) input power, the piezoelectric field induced by the SAWs significantly modifies the material's intrinsic band structure. The SAW in this work is Rayleigh-mode, featuring both in-plane ($E_x$) and out-of-plane ($E_z$) piezoelectric field components. These components can be utilized to modulate intralayer excitons with large in-plane polarizability and interlayer excitons with out-of-plane dipole moments, respectively. The corresponding three-dimensional finite element simulation results are presented in Figure 1c, Supplementary Note 5 and the Supplementary Figure 4, which demonstrate the electric field distribution at the $LiNbO_3$ surface along the SAW propagation path. Figure 1d shows the simulated electric potential along a line section at the $LiNbO_3$ surface in the SAW propagation direction, revealing that the in-plane piezoelectric field component is approximately 2.07 times stronger than the out-of-plane component, RF power used in simulations is shown in Supplementary Note 6 and Supplementary Figure 5. The intimate contact between the monolayer material and the piezoelectric substrate promotes type-II band edge modulation, resulting in the formation of periodic potential wells with a spatial separation of the SAW wavelength. While SAWs inherently generate both piezoelectric electric fields and mechanical strain, our analysis indicates that the contribution of dynamic mechanical strain to exciton modulation is negligible compared to the dominant piezoelectric electric field effect (see Supplementary Note 7 for detailed quantitative analysis). Therefore, our subsequent discussion focuses primarily on the electro-acoustic modulation mechanisms. Figure 1e provides a simplified schematic of this modulation mechanism. The strong piezoelectric field not only separates photoexcited excitons but also confines electrons and holes at the CBM and VBM with a drift direction along with the SAW, providing a novel route to manipulate the excitons in vdW heterostructures.

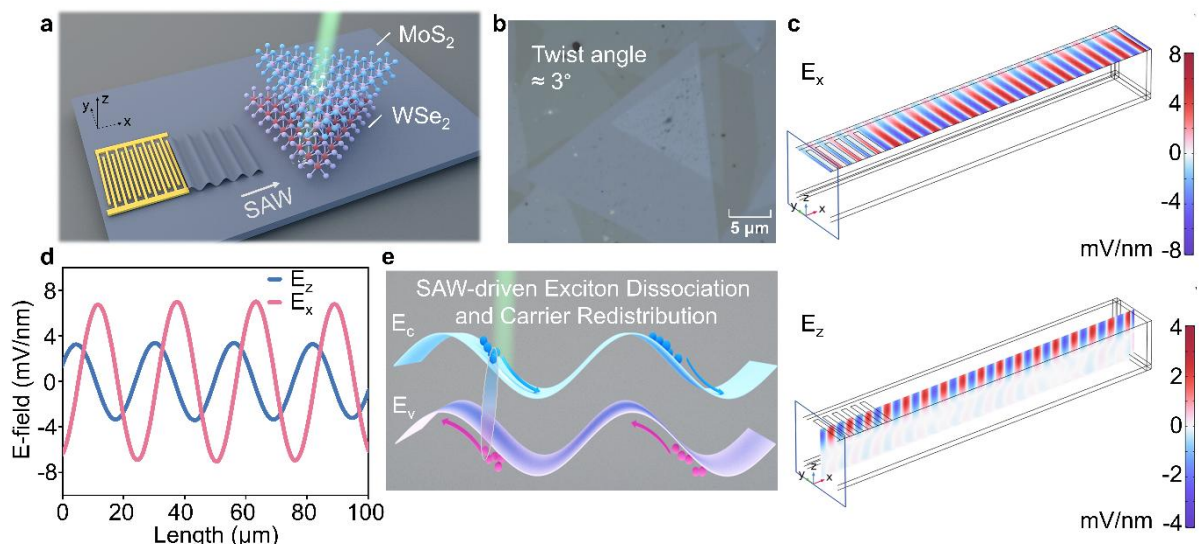

**Figure 1. $MoS_2$/$WSe_2$ heterostructure integrated with SAW devices. a,** Schematic of the SAW modulation platform with a $MoS_2$/$WSe_2$ heterostructure stacked on the path of the SAW propagation. Interdigital transducers (IDTs) generate SAWs that propagate along the indicated direction. **b,** Optical microscope image of the $MoS_2$/$WSe_2$ heterostructure. The twist angle is estimated to be approximately 3°. **c,** The in-plane piezoelectric field components ($E_x$) and out-of-plane piezoelectric field components ($E_z$) obtained from 3D finite element simulations. **d,** Simulated piezoelectric field along a line section on the $LiNbO_3$ surface in the SAW propagation direction, demonstrating that $E_x$ is approximately 2.07 times that of $E_z$. **e,** Schematic of SAW-induced type-II band modulation, exciton dissociation, and carrier redistribution.

**Figure 2** displays the measured PL spectra of excitonic emissions from the $WSe_2$ and $MoS_2$ monolayers and the twisted heterostructure region at 4 K. Figure 2a shows the neutral exciton emission of $MoS_2$ ($X^0_{MoS_2}$), while Figure 2b displays the fitting results of multiple excitonic emissions in $WSe_2$, where the best-fit spectrum (red translucent line) closely matches the raw data. The contributions of individual excitonic components are represented by lines of different colors. The highest-energy component (pink), located at approximately 1.76 eV, is attributed to the neutral bright exciton ($X^0_{WSe_2}$). The purple line represents the negatively charged excitons ($X^-_{WSe_2}$), and the orange line corresponds to the composite emission of biexcitons ($XX_{WSe_2}$) and charged biexcitons ($XX^-_{WSe_2}$). The blue and the green spectra are likely attributed to the localized exciton (LX) and low-energy tail (see Supplementary Note 8), respectively[30-34]. A comprehensive discussion of the assignments of these excitonic spectral lines is provided in Supplementary Note 9. As shown in Figure 2c, the excitons from the $WSe_2$ and $MoS_2$ monolayers remain visible in the heterostructure region. However, due to the changes in the dielectric environment at the interface, their energies exhibit slight shifts, and their intensities are significantly quenched, which can be attributed to the rapid charge transfer during the formation of interlayer excitons. In addition, two new peaks are observed near 1.65 eV and 1.67 eV in the heterostructure region, which are absent in the monolayers. These peaks could be assigned to indirect neutral excitons $X^0_I$ and indirect negatively charged trions $X^-_I$ formed by electrons in the K valley of $WSe_2$ and holes in the hybridized Γ valley of the heterostructure[35-37]. Additionally, we observe that the $X^-_I$/$X^0_I$ intensity ratio decreases with increasing temperature and eventually saturates. The binding energy of the $X^-_I$ is 28 meV. This experimental trend is consistent with the mass action law for indirect trions[38], as shown in Figure 2d (detailed derivations in Supplementary Note 10), offering clear evidence for assigning the two spatially indirect emission lines to neutral excitons and charged trions, respectively, and demonstrating that trions are more susceptible to thermal dissociation at higher temperatures.

Within the near short-wave infrared (nSWIR) range (Figure 2e), we detect excitonic splitting at energies of 0.95 eV and 0.83 eV. These splittings are attributed to interlayer neutral excitons $IX_{K-\Gamma}$ in the K-Γ valley and interlayer excitons $IX_{K-K}$ in the K-K valley. In contrast to previous experimental results, which reported only one K-K interlayer exciton in this range[25,26],[39], our experiments reveal two interlayer excitons with a significant energy splitting of 120 meV, induced from different valley states. In contrast to observations reported in previous studies[25-27], our experiments reveal a significant redshift in the exciton peak energy, see Supplementary Note 11 and Supplementary Figure 6. Notably, this splitting remains stable across all twist-angle samples in the range of 3 to 50 degrees (Supplementary Notes 12 and Figure 7). We also acknowledge that the absolute peak positions of both $IX_{K-\Gamma}$ and $IX_{K-K}$ exhibit twist-angle dependence, consistent with observations in previous experimental studies[26]. Furthermore, the intensity ratio between $IX_{K-\Gamma}$ and $IX_{K-K}$ also changes with twist angle (Supplementary Figure 9). This twist-angle-dependent property is consistent with the trend predicted by DFT calculations (Supplementary Figure 10). We have summarized the fully labeled band diagram for the $MoS_2/WSe_2$ heterostructure to provide a clearer illustration of the various types of excitons arising from interlayer and intralayer transitions between the K–K and K–Γ valleys (Supplementary Figure 11).

Detailed first-principles calculations are performed to identify the types of interlayer excitons. An AB-stacked vdW $WSe_2/MoS_2$ Moiré superlattice with a twist angle of 3.11 degrees is constructed (Figure 2f). Theoretical calculations show that the Moiré superlattice reveals quasi-direct bandgap properties of monolayers, with the CBM and VBM at the K-point in the Brillouin zone, contributed by the $MoS_2$ and $WSe_2$ layers, respectively (Figure 2g). This interlayer transition with an energy of 1.0 eV at GGA level can be assigned to $IX_{K-K}$, and the electron transfer effect from $WSe_2$ to $MoS_2$ is observed. An indirect transition $IX_{K-\Gamma}$ with an energy 58 meV higher than the K-K transition is further revealed, which derives from the orbital hybridization at the Γ point due to the strong interlayer coupling between $MoS_2$ and $WSe_2$. To correct the bandgap underestimation of GGA, the HSE06 hybrid functional was used. Since HSE06 calculations are limited to models with fewer than 100 atoms, we adopted a small model for the correction. As shown in Supplementary Figure 7, the energy difference between the two excitons remains comparable across models with different twist angles, confirming the validity of this approach. The HSE06 result yields an energy difference ($E_{K-\Gamma}$-$E_{K-K}$) of 121 meV (Supplementary Figure 8), which is in agreement with the experimental value.

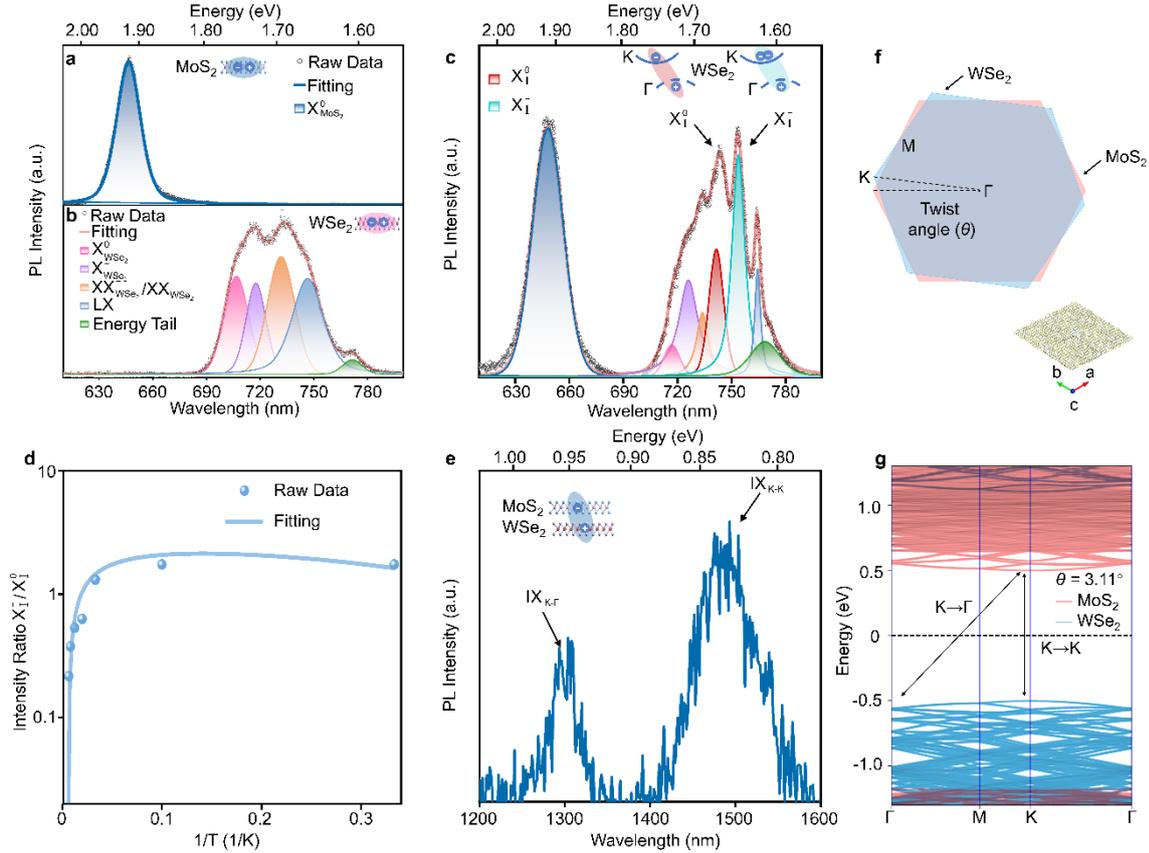

**Figure 2. PL Spectra and Exciton calculations in the Heterostructure. a, b,** Measured PL spectra from the monolayer regions of MoS₂ and WSe₂ at 4 K, respectively. **c,** Measured PL spectra from the MoS₂/WSe₂ heterostructure region at 4 K. Peak splitting induced by indirect neutral excitons $X_I^0$ and indirect negatively charged trions $X_I^-$ are labeled. **d,** Dependence of the $IX_{K-\Gamma}^-/IX_{K-\Gamma}^0$ intensity ratio on the temperature. **e,** Measured PL spectra from the MoS₂/WSe₂ heterostructure region at 4 K. The Peak splitting induced by interlayer neutral excitons in K-Γ valleys and K-K valleys in the nSWIR range are labeled. **f,** AB-stacked vdW MoS₂/WSe₂ Moiré superlattice with a twist angle of 3.11°. **g,** Calculated electronic band structure of the MoS₂/WSe₂ Moiré superlattice, showing a quasi-direct bandgap with the CBM and VBM located at the K-point in the Brillouin zone, originating from the MoS₂ and WSe₂ layers, respectively.

## 2.2. Effects of SAW and Temperature on Excitons in the VIS and NIR Range of Heterostructures

We further explore the modulation of excitonic PL intensity by SAWs and temperature. First, we observe two opposing effects for the excitons in the visible to near-infrared range: PL intensity enhancement and quenching. **Figure 3**a-c depicts the evolution of PL spectra as the SAW power increases from 0 to 15 dBm at 4 K. Figure 3d and e summarize the variation of exciton PL intensity with SAW power. The emission intensity of the $X^0$ excitons in WSe$_2$ and MoS$_2$ shows a slight enhancement with increasing SAW power, which can be explained by the Poole-Frenkel effect[40,41] at low temperatures. Through this mechanism, SAWs reduce trap potential barriers in two-dimensional materials, decrease bound carrier activation energy, and promote electron-hole transitions to bound states in the continuum, thereby enhancing exciton formation and increasing PL signal intensity. For the indirect neutral excitons $X_I^0$ and indirect negatively charged trions $X_I^-$ in the VIS and NIR range, the PL emission is suppressed upon SAW activation, with the suppression increasing steadily with SAW power. This SAW-induced quenching of excitonic PL in two-dimensional materials has been widely observed[14],[18-22]. The suppression in PL intensity arises from the interaction between the SAW piezoelectric field and photoexcited excitons. The spatially modulated SAW piezoelectric potential induces type-II band modulation, leading to lateral separation of electrons and holes, extended carrier lifetimes, and enhanced non-radiative recombination rates[14],[18]. Furthermore, some carriers are trapped at the maxima and minima of the propagating piezoelectric potential and are transported away from the laser excitation spot, further contributing to the PL suppression, as shown in Figure 1e. Figure 3f shows the heatmap of PL spectra of the heterostructure region in the 725-780 nm wavelength range. As the SAW power increases from 0 to 25 dBm, a significant PL quenching effect is observed. The differential response of neutral and indirect exciton PL emission to SAW modulation is primarily attributed to the significantly larger binding energy of neutral excitons.

Subsequently, we examine the impact of temperature on excitonic PL. As illustrated in Figure 3g, the PL intensity exhibits a pronounced enhancement with decreasing temperature. This behavior can be attributed to two primary mechanisms: first, the low temperature leads to a reduction of the momentum mismatch between band extrema[42], thereby enhancing the exciton recombination probability; second, the reduction in temperature leads to a significant decrease in non-radiative recombination channels, further facilitating the radiative recombination process of excitons[43]. An increase in temperature leads to a red shift in the exciton peak position. The thermal expansion of the material lattice at higher temperatures narrows the band gap, leading to a reduction in the exciton emission energy and consequently causing a red shift[44]. This observation aligns with the Varshni equation, which describes the temperature-dependent bandgap variation in conventional semiconductors. The Varshni equation is expressed as: $E_g(T) = E_0 - \alpha T^2/(T + \beta)$, where $E_g(T)$ represents the material's bandgap at temperature $T$, $E_0$ denotes the bandgap at 0 K, and $\alpha$ and $\beta$ are fitting parameters[45]. With increasing temperature, the splitting of indirect excitons vanishes at around 180 K, in agreement with the mass action law depicted in Figure 2d. This quenching effect is primarily attributed to the enhanced thermal dissociation of $X_I^-$ with lower binding energy at higher temperatures[46]. After establishing the temperature-dependent evolution of excitonic PL, we subsequently investigate the effect of SAW modulation at room temperature (297 K). Figure 3h presents the PL spectrum under a SAW power of 15 dBm at 297 K (the spectra without SAW modulation and under other SAW power levels are provided in Supplementary Note 13 and Figure 12). The exciton peaks at 297 K exhibit a significant broadening effect compared with the measured spectra at 4 K. This thermal broadening is attributed to enhanced lattice vibration (phonon scattering) at elevated temperatures, which boosts the exciton-phonon interaction[47]. Upon SAW excitation, $X_{WSe_2}^0$, $X_{MoS2}^0$, and $X_I^0$ all three broad exciton peaks demonstrate pronounced quenching effects as the SAW power increases, as depicted in Figure 3i. The thermal map illustrating the modulation of excitons $X_{WSe2}^0$ and $X_I^0$ by SAW at 297 K is summarized in Supplementary Note 14 and Figure 13. The variation of PL with SAW power at other temperatures is shown in Supplementary Note 15 and Figure 14.

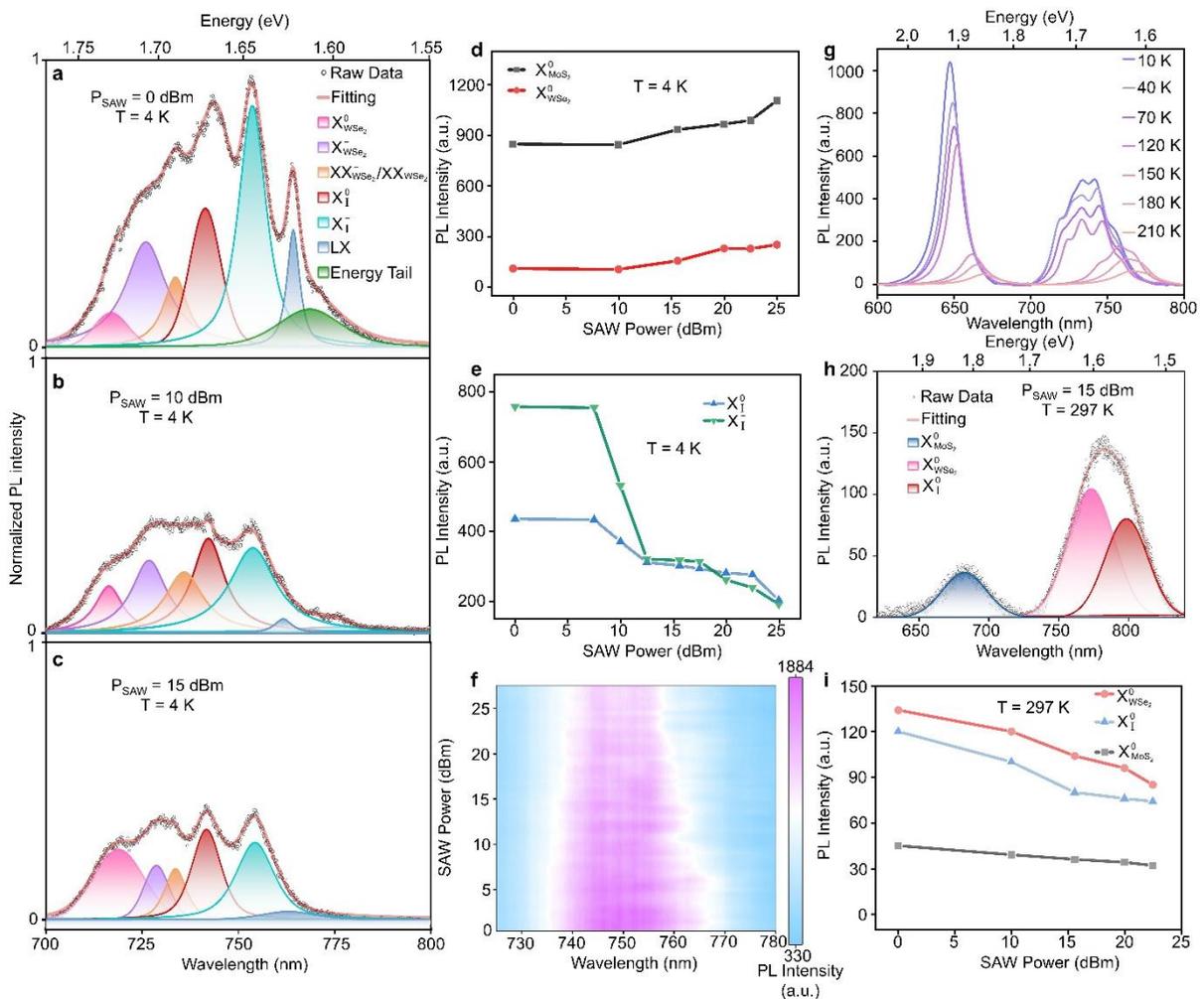

**Figure 3. SAW Induced Exciton PL Quenching and Temperature Dependent Exciton Modulation in the Visible/NIR range. a ~ c,** Measured PL spectra of SAW-modulated excitons under varying excitation power at 4 K. **d,** PL intensity of the $X_{WSe2}^0$ and $X_{MoS2}^0$ excitons exhibits a

slight enhancement with increasing SAW power. **e,** Collected PL intensity of the $X_I^0$ and $X_I^-$ excitons in the heterostructure region decreases with increasing SAW power. **f,** PL intensity of $X_I^0$ and $X_I^-$ as a function of SAW excitation power is shown in the heatmap, at 4K. **g,** PL spectra measured at varying temperatures in the heterostructure region. **h,** The PL spectrum measured at 297 K under a 15 dBm SAW modulation across the visible to near-infrared range. **i,** The PL intensities of $X_{WSe2}^0$, $X_{MoS2}^0$ and $X_I^0$ measured as a function of SAW power at 297 K.

## 2.3. Effects of SAW and Temperature on Excitons in the nSWIR Range of Heterostructures

We systematically investigate the effects of temperature and SAW modulation on interlayer excitons in the nSWIR range. **Figure 4a** presents the PL spectra of interlayer excitons at different temperatures, processed by Gauss–Lorentz peak fitting to reduce noise and extract reliable peak parameters. As the temperature increases from 4 K to 100 K, the intensities of both exciton peaks decrease gradually, and the peaks vanish at temperatures above 100 K. Meanwhile, a comparison of the peak widths reveals that the $IX_{K-K}$ peak is about twice as broad as the $IX_{K-\Gamma}$ peak. This broadening is mainly attributed to stronger repulsive interactions between their electric dipoles[47], as interlayer excitons, formed by electrons and holes in spatially separated monolayers, inherently possess a vertical electric dipole moment. The repulsive force between excitons is given by $F = -\frac{dU}{dr} \approx -(\frac{3p^2}{4\pi\varepsilon r^4})$, where $p$ is the dipole moment, $r$ is the separation, and $\varepsilon$ is the dielectric constant. Thus, the significant broadening of the $IX_{K-K}$ spectrum indicates a larger dipole moment and stronger exciton-exciton repulsion. Figure 4b plots the ratio $IX_{K-K}/IX_{K-\Gamma}$ as a function of temperature. The ratio decreases with increasing temperature, indicating that $IX_{K-K}$ has a lower binding energy and is more susceptible to thermal dissociation than $IX_{K-\Gamma}$. In Figure 4c, we examine the dependence of the ratio of the two excitons with SAW power at 4 K. The results show that $IX_{K-K}/IX_{K-\Gamma}$ decreases as SAW power increases, signifying that $IX_{K-K}$ is more sensitive to the SAW-induced acoustoelectric field. This stronger response is primarily due to its larger electric dipole moment. The electric field modulates the exciton energy via $\Delta U = -p \cdot E$, making $IX_{K-K}$ more prone to dissociation into free electrons and holes, which can subsequently travel with the propagating SAW. Figure 4d shows the changes in the PL spectrum of interlayer excitons at 70 K under different SAW powers, clearly indicating that higher SAW power leads to stronger quenching of the PL.

To quantitatively analyze the quenching effect, Figure 4e presents the quenching index of the K-K interlayer exciton as a function of SAW power at different temperatures. The quenching index is defined as $Q_{PL} = 1 - \frac{I_{IX_{K-K}}(\text{with SAW})}{I_{IX_{K-K}}(\text{without SAW})}$. The results demonstrate that at 4 K, 18 dBm SAW excitation quenches the IX$_{K-K}$ PL by only ~10%. However, as the temperature increases, the quenching efficiency is significantly enhanced, and the quenching index approaches 1 for 18 dBm at 70 K, indicating near-complete quenching. This temperature-enhanced SAW-induced quenching is mainly attributed to two mechanisms. First, at higher temperatures, thermal dissociation of excitons significantly increases the free carrier density, thereby amplifying the overall exciton-SAW interactions (Supplementary Note 16). Second, the increased temperature promotes exciton diffusion, which further improves the exciton response to moving SAWs[48].

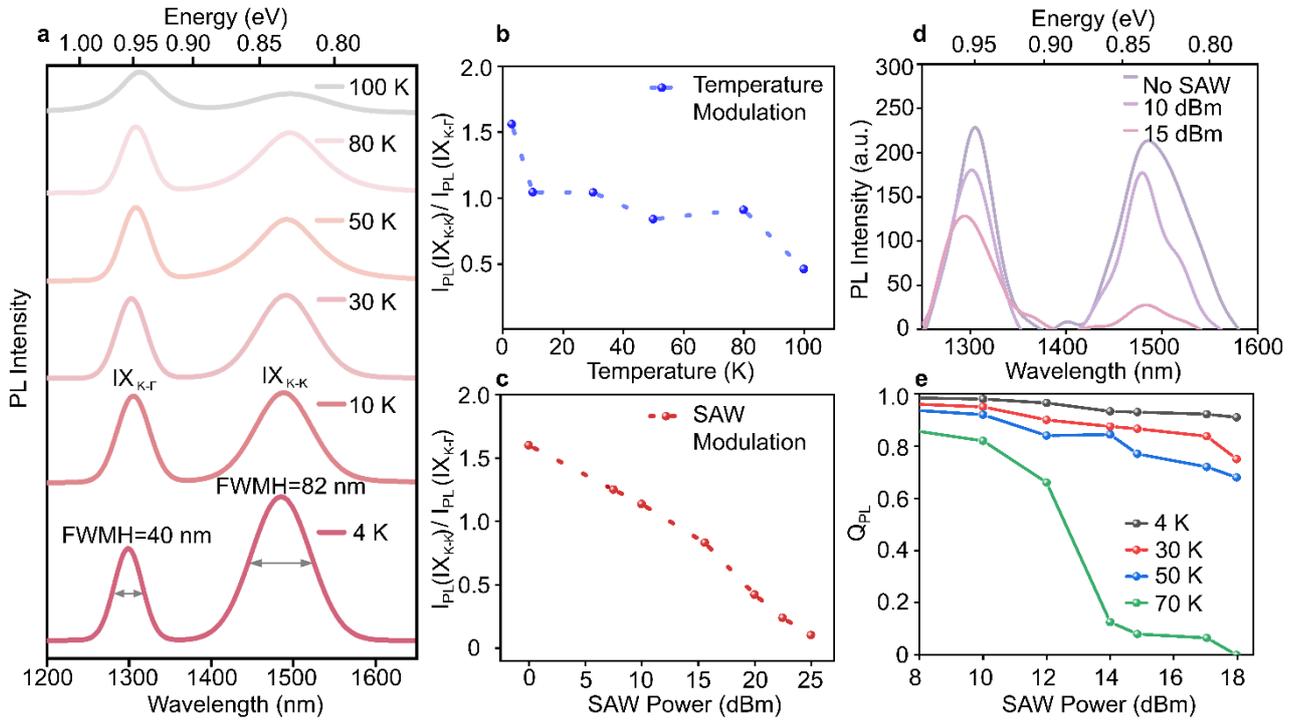

**Figure 4. SAW-Induced Exciton PL Quenching and Thermal Exciton Modulation in the nSWIR Range. a,** Temperature-dependent exciton PL spectra of IX$_{K-K}$ and IX$_{K-Γ}$. **b, c,** Proportion of IX$_{K-K}$ and IX$_{K-Γ}$ as a function of temperature and SAW power, respectively. **d,** Quenching of K-Γ and K-K interlayer exciton emission spectra under SAW excitation at 70 K. **e,** Quenching index of the K-K interlayer exciton as a function of SAW power at different temperatures, showing enhanced quenching at higher temperatures.

### 2.4. The Stark effect of Excitons Modulated by SAW

In the investigation of the excitonic PL energy in the MoS$_2$/WSe$_2$ heterostructure region as a function of SAW electric field, we observe two distinct Stark effects: a linear Stark effect for interlayer excitons[49] and a quadratic Stark effect for intralayer excitons[14],[21] (the detailed derivation of this effect is provided in Supplementary Note 17). As illustrated in **Figure 5**a, Gaussian fitting of the PL spectra allows us to extract the energy of interlayer K-K valley and K-Γ valley excitons as a function of the out-of-plane SAW electric field $E_z$, in the nSWIR range. Interlayer excitons, formed by spatially separated electrons and holes in different layers, exhibit a permanent out-of-plane electric dipole moment. The out-of-plane polarizability significantly surpasses the in-plane polarizability, resulting in a linear Stark effect in their energy as modulated by the out-of-plane component of the piezoelectric field. The energy shift is described by $\Delta E = -p \cdot E_z$, where $p$ is the electric dipole moment and $E_z$ is the applied SAW electric field. The electric dipole moment of interlayer excitons points from the negative charge (located in the $MoS_2$ layer) to the positive charge (located in the $WSe_2$ layer). As shown in Figure 5a, although the alternating polarity of the SAW electric field is expected to induce both blueshifts and redshifts in the interlayer exciton energy, our experimental findings show only a blueshift. When the SAW electric field is anti-parallel to the exciton dipole, it not only increases the exciton energy (blueshift), but also spatially confines electrons and holes, thereby increasing their wavefunction overlap and promoting radiative recombination[36],[50]. In contrast, when the field is parallel to the dipole (redshift phase), electron and hole wavefunctions are further separated, leading to efficient exciton dissociation and non-radiative recombination, which suppresses the PL signal. As a result, the detected PL selectively originates from blueshifted, radiatively recombining excitons. Detailed explanation is provided in Supplementary Note 18 and Supplementary Figure 15. By fitting the slope of the curve in Figure 5a, we determine the dipole moment of the K-K interlayer exciton to be ~ 0.86 e·nm. This pronounced field-tuning behavior is inconsistent with emission from states confined to a single layer, directly confirming the interlayer nature of this state. Notably, the slope of the linear Stark shift for the $IX_{K-K}$ exciton observed in our SAW-based measurement (~0.86 e·nm) is remarkably close to that reported by Karni et al. using a static vertical electric field (0.5~0.8 e·nm) [25]. This agreement suggests that the SAW-induced piezoelectric field in our platform can achieve an effective field strength comparable to conventional static-gate configurations, thereby validating the quantitative accuracy of the SAW approach. Minor deviations may arise from the high dielectric constant environment of the $LiNbO_3$ substrate[51]. Meanwhile, the fitted dipole moment of the K-Γ interlayer exciton is found to be 0.72 e·nm. This relatively smaller value indicates that it likely represents a hybrid exciton formed by the combination of interlayer and intralayer states. To validate this point, we further perform DFT calculations for the spatial distribution of interlayer electrons and holes, and analyze the orbital hybridization effect between $WSe_2$ and $MoS_2$ by projected density of states (Figure 5b). The electronic states of $MoS_2$, including the 3$d$ orbitals of Mo and the $2p$ orbitals of S atoms, undergo a strong band hybridization with the valence band of WSe$_2$ in the range of ~ 0.1-0.7 eV below the VBM of the heterostructure, corresponding to the hole hybridized states in the Γ valley (Figure 2g), denoted as $|+\Gamma\rangle$. Three unique Bloch states involved in two interlayer transitions (K-K and K-Γ, respectively) are further visualized in Figure 5c. The electron $|-\rangle$ and hole $|+K\rangle$ states of the K-valley are entirely contributed by individual MoS$_2$ and WSe$_2$ layers, respectively. While the Γ-valley hole state $|+\Gamma\rangle$ is hybridized by the electronic states of MoS$_2$ and WSe$_2$ with occupancies of about 28% and 72%, respectively, which represent the intra- and interlayer transition intensities of the K-Γ transitions. Three-dimensional isosurface mapping of the Bloch wavefunction further reveals strong interlayer coupling at the interface, with both the $d$ orbitals of the transition metal atoms and the $p$ orbitals of the sulfur atoms involved in band hybridization.

For intralayer excitons, we observe the quadratic Stark effect. Since intralayer excitons lack a permanent dipole moment, the SAW-induced energy shift of the exciton peak is primarily driven by polarization induced by the piezoelectric field. Given the significantly higher in-plane polarizability of intralayer excitons compared with the out-of-plane polarizability (with the out-of-plane component of the SAW field ($E_z$) being approximately half of the in-plane component ($E_x$), see Figure 1d), we exclusively consider the second-order Stark shift induced by the in-plane piezoelectric field ($E_x$) and the in-plane polarizability, described by $\Delta E = -\frac{1}{2}\alpha E_x^2$. As shown in Figure 5 d, we determine the in-plane polarizability of the X$_0$ exciton in WSe$_2$ to be $\alpha_{\text{WSe}_2}^0 \approx (550 \pm 30) \times 10^{-5}$ meV/(kV/cm)$^2$. This value is significantly higher than the polarizability of excitons in h-BN encapsulated WSe$_2$ monolayers ($\alpha_{\text{WSe}_2}^0 \approx 10.4 \times 10^{-5}$ meV/(kV/cm)$^2$)[52], primarily due to the reduced exciton binding energy in a high dielectric environment (Supplementary Note 19).

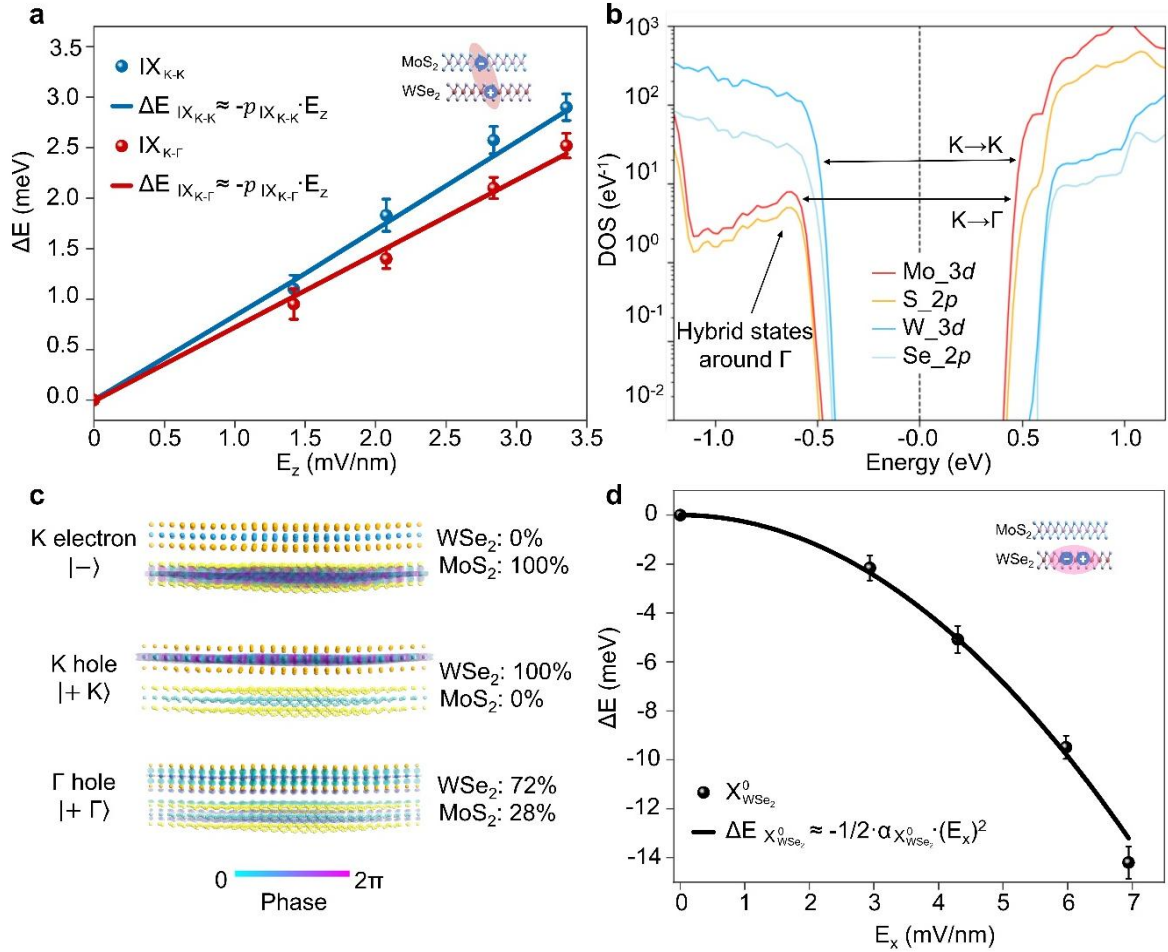

**Figure 5. SAW-induced Stark effect for intralayer and interlayer excitons. a,** The linear Stark effect of interlayer exciton, due to the coupling between $E_z$ and the exciton's out-of-plane dipole moment. **b,** Orbital hybridization effect between $WSe_2$ and $MoS_2$ analyzed by projected density of states. **c,** Three Bloch states associated with K-K and K-Γ interlayer transitions. The isosurface of the wavefunction is 0.02 Å$^{-1.5}$. **d,** The energy of intralayer exciton modulated by the quadratic Stark effect, which results from the interaction between the e exciton's in-plane polarizability and the in-plane electric field ($E_x$) of the SAW.

## 3. Conclusion

In summary, this work demonstrates a platform that enables SAW modulation of both intralayer and interlayer excitons in $MoS_2/WSe_2$ heterostructures. By combining DFT calculations and experimental observations, we identify different types of excitons, including indirect excitons within the VIS and NIR region and low-energy interlayer excitons in both the K-K and K-Γ valleys in the nSWIR regime, thereby enriching the diversity of interlayer excitonic species. By utilizing both in-plane and out-of-plane piezoelectric fields generated by SAWs, we achieve precise and continuous tuning of exciton energy levels via Stark shifts, with modulation depths an order of magnitude smaller than exciton linewidths. This also enabled effective modulation of exciton PL spectra via SAW-induced exciton dissociation, all achieved at a fixed center resonant SAW frequency. The results reveal that SAWs can selectively manipulate interlayer and intralayer exciton emission properties without direct electrical contact, thus offering a versatile platform for exciton routing and control in TMD heterostructures. The observed stable valley-specific exciton splitting, especially the O-band and S-band emissions from nSWIR interlayer excitons, demonstrates promising potential for on-chip excitonic light sources. Collectively, these findings advance our understanding of SAW-exciton interactions and lay the groundwork for the development of next-generation optoelectronic and quantum devices based on exciton engineering.

## 4. Experimental Section/Methods

***Sample Fabrication***: On a lithium niobate (LN) substrate, we fabricated a series of acoustic delay lines with interdigital transducers (IDTs) using ultraviolet lithography. The IDT is designed with each IDT comprising exactly 120 finger pairs and a 1:1 duty cycle, optimized to efficiently excite and receive surface acoustic waves at a wavelength of 25 μm, which corresponds to a design frequency of approximately 160 MHz at 297 K. The TMD monolayers were prepared via a two-step process: first, high-quality $MoS_2$ and $WSe_2$ monolayers were grown on sapphire substrates using the chemical vapor deposition (CVD) method (supplied by Sixcarbon Technology Shenzhen). Subsequently, to construct $MoS_2$/$WSe_2$ heterostructures with specific twist angles, we utilized a transfer technique based on poly(methyl methacrylate) (PMMA). Following each monolayer transfer onto the LN substrate, an annealing process was conducted at 150°C for one hour in a vacuum environment to remove PMMA residues and enhance the contact between the monolayers and the substrate. Notably, various twist angles were randomly formed between the $MoS_2$ and $WSe_2$ monolayers, significantly influencing the heterostructure properties. Consequently, we meticulously selected samples with different twist angles for systematic investigation and characterization.

***Electrical characterization and SAW excitation***

The scattering parameters $S_{11}$ (reflection) and $S_{21}$ (transmission and insertion loss) of the IDTs were assessed using a vector network analyzer (Rohde & Schwarz ZNA67) to measure the scattering parameters of the RF network. The RF excitation of SAW is applied via a signal generator (AFG31252, Tektronix) and further amplified by an amplifier to enhance the SAW power. At each temperature, we swept the RF frequency around resonance at fixed input power to locate the center resonance frequency, identified as the point of maximum PL quenching, thereby ensuring optimal RF-to-acoustic conversion efficiency. In the final measurements, the RF excitation was applied at the fundamental center resonance frequency, and single-point frequency excitation was performed for data acquisition (Supplementary Note 20 and Figure 16).

***Acoustic Modulation µ-PL Spectroscopy Measurements***: The temperature-dependent PL measurements were performed using a liquid helium circulation cryogenic system, covering a temperature range of 4 K to 297 K. To reduce possible heating effects, the integration time was 1 s for weaker nSWIR signals and 0.5 s for stronger VIS/NIR signals, with 1 min cooling intervals between successive acquisitions. Localized heating under SAW excitation has no observable effect on the PL measurements (see Supplementary Note 21 and Figure 17). This low-temperature system exhibits minimal vibration, making it ideal for high-sensitivity microscopic optical measurements. A continuous-wave laser with a wavelength of 532 nm was employed as the excitation source, and the laser was focused onto the sample surface using a 20× objective lens (NA = 0.40) to create a spot with a measured diameter of 2.6 µm. The spot size at the sample plane was determined using a direct imaging method in which the focused beam profile was imaged onto a calibrated CCD camera. Spectral signals in the visible range were acquired using a Horiba spectrometer, while infrared spectral signals were collected with an Oceanhood NIRPro near-infrared spectrometer coupled to a liquid helium-cooled CCD camera. The SAW device is excited using a signal generator. The SAW device is wire-bonded to a custom PCB with aluminum wires to enable PL measurements in the cryogenic chamber. Concurrently, electrical excitation is applied.

***Finite Element Method Simulation***: We employed the finite element method (COMSOL Multiphysics 6.2) to construct a three-dimensional model for calculating the piezoelectric field strength generated by IDT resonance on $LiNbO_3$ substrates. The model successfully simulated both the in-plane (Ex) and out-of-plane (Ez) electric field splitting under different excitation powers, as well as the radio-frequency scattering (S) parameters of the device. The simulated resonance frequency closely matched the experimentally measured value. Detailed parameter settings and simulation results are provided in the Supplementary Information.

***DFT Calculations***: All first-principles calculations were performed by using the QuantumATK package (version V-2023.12)[53]. The exchange-correlation functional was based on Perdew-Burke-Ernzerho (PBE) generalized gradient approximation (GGA)[54]. The interactions between valence electrons and ion cores were described by the SG15 optimized norm-conserving Vanderbilt (ONCV) pseudopotentials[55]. The quasi-particle approximation DFT-1/2 approach[56,57] and the Heyd-Scuseria-Ernzerhof (HSE06) hybridization functional[58,59] were also considered in the energy bands calculation to correct the bandgap underestimation derived from the GGA-PBE calculation. We used the Grimme DFT-D3 scheme to describe the weak van der Waals interactions[60]. To avoid periodic mirror pseudo-interactions, the vacuum layer was set to 20 Å and the dipole correction was applied along the surface-normal $z$-direction. The maximum residual force on each atom was set to be less than 0.02 eV Å$^{-1}$, and the self-consistent field convergence tolerance was set to $10^{-5}$ during structure relaxation. The real-space mesh cutoff was chosen to be 180 Hartree and the electron temperature was set to 300 K for Fermi-Dirac broadening. An AB stacked van der Waals Moiré superlattice with a twist angle of 3.11 degrees was constructed, which contained a $\sqrt{217} \times \sqrt{217}$ monolayer $MoS_2$ and a $\sqrt{199} \times \sqrt{199}$ monolayer $WSe_2$ supercell with strains of - 0.04% and 0.04%, respectively, for a total of 1,248 atoms. The Brillouin zone was integrated only by sampling the Gamma point due to the large lattice constant of 46.89 Å of the supercell.

## Supporting Information

Supporting Information is available from the Wiley Online Library or from the author.

## Acknowledgements

This work was supported by the National Natural Science Foundation of China under Grant No. 62574084 (CZ), and Guangdong Provincial Key Field Research and Development Program 2022B0701180002 (CZ). We acknowledge the support from the Micro & Nano Electronics Platform (MNEP) of South China University of Technology (SCUT) for device fabrication and characterization.

## Data Availability Statement

All data needed to evaluate the conclusions in the paper are present in the paper and/or the Supplementary Materials. Additional data related to this paper may be requested from the authors.

## Competing interests

Authors declare that they have no competing interests.

**References**

[1] Regan, E. C., Wang, D., Paik, E. Y. et al. Emerging exciton physics in transition metal dichalcogenide heterobilayers. *Nature Reviews Materials* **7**, 778–795 (2022)

[2] Lian, Z., Meng, Y., Ma, L. et al. Valley-polarized excitonic Mott insulator in $WS_2/WSe_2$ moiré superlattice. *Nature Physics* **20**, 34–39 (2024).

[3] Tan, Q., Rasmita, A., Zhang, Z. et al. Enhanced coherence from correlated states in $WSe_2/MoS_2$ moiré heterobilayer. *Nature Communications* **16**, 1–7 (2025).

[4] He, K., Kumar, N., Zhao, L. et al. Tightly bound excitons in monolayer $WSe_2$. *Physical Review Letters* **113**, 026803 (2014).

[5] Gao, K., Li, Y., Yang, Y. et al. Manipulating coherent exciton dynamics in $CsPbI_3$ perovskite quantum dots using magnetic field. *Advanced Materials* **36**, 2309420 (2024).

[6] Goryca, M., Li, J., Stier, A. V. et al. Revealing exciton masses and dielectric properties of monolayer semiconductors with high magnetic fields. *Nature Communications* **10**, 4172 (2019).

[7] Li, H., Chen, F., Jia, H. et al. All-optical temporal logic gates in localized exciton polaritons. *Nature Photonics* **18**, 864–869 (2024).

[8] Pimenta Martins, L. G., Ruiz-Tijerina, D. A., Occhialini, C. A. et al. Pressure tuning of minibands in $MoS_2/WSe_2$ heterostructures revealed by moiré phonons. *Nature Nanotechnology* **18**, 1147-1153 (2023).

[9] Kim, J. M., Jeong, K. Y., Kwon, S. et al. Strained two-dimensional tungsten diselenide for mechanically tunable exciton transport. *Nature Communications* **15**, 10847 (2024).

[10] Xu, H., Wang, J., Liu, H. et al. Control of hybrid exciton lifetime in $MoSe_2/WS_2$ moiré heterostructures. *Advanced Science* **11**, 2403127 (2024).

[11] Janker, L., Tong, Y., Polavarapu, L., Feldmann, J., Urban, A. S., and Krenner, H. J. Real-Time Electron and Hole Transport Dynamics in Halide Perovskite Nanowires. *Nano Letters* **19**, 8701-8707 (2019).

[12] Lienhart, M., Choquer, M., Nysten, E. D. et al. Heterogeneous integration of superconducting thin films and epitaxial semiconductor heterostructures with lithium niobate. *Journal of Physics D: Applied Physics* **56**, 365105 (2023).

[13] Sonner, M. M., Gnedel, M., Berlin, J. C., Rudolph, D., Koblmüller, G., and Krenner, H. J. Sub-nanosecond acousto-electric carrier redistribution dynamics and transport in polytypic GaAs nanowires. *Nanotechnology* **32**, 505209 (2021).

[14] Datta, K., Li, Z., Lyu, Z. and Deotare, P. B. Piezoelectric Modulation of Excitonic Properties in Monolayer $WSe_2$ under Strong Dielectric Screening. *ACS Nano* **15**, 12334-12341 (2021).

[15] Hernández-Mínguez, A., Liou, Y. & Santos, P. Interaction of surface acoustic waves with electronic excitations in graphene. *Journal of Physics D: Applied Physics* **51**, 383001 (2018).

[16] Qi, R., Joe, A. Y., Zhang, Z. et al. Perfect Coulomb drag and exciton transport in an excitonic insulator. *Science* **388**, 278–283 (2025).

[17] Nie, X., Wu, X., Wang, Y. et al. Surface acoustic wave induced phenomena in two-dimensional materials. *Nanoscale Horizons* **8**, 158-175 (2023).

[18] Gomes, M. L. F., Matrone, P. W., Cadore, A. R., Santos, P. V. and Couto, O. D. D. Acoustic Modulation of Excitonic Complexes in $hBN/WSe_2/hBN$ Heterostructures. *Nano Letters* **24**, 15517-15524 (2024).

[19] Polimeno, L., Di Renzo, A., Rizzato, S. et al. Dynamic Mechanical Modulation of $WS_2$ Monolayer by Standing Surface Acoustic Waves. *ACS Photonics* **11**, 4058-4064 (2024).

[20] Rezk, A. R., Carey, B., Chrimes, A. F. et al. Acoustically-Driven Trion and Exciton Modulation in Piezoelectric Two-Dimensional $MoS_2$. *Nano Letters* **16**, 849-855 (2016).

[21] Scolfaro, D., Finamor, M., Trinchao, L. O. et al. Acoustically Driven Stark Effect in Transition Metal Dichalcogenide Monolayers. *ACS Nano* **15**, 15371-15380 (2021).

[22] Sheng, L., Tai, G., Yin, Y., Hou, C. and Wu, Z. Layer-Dependent Exciton Modulation Characteristics of 2D $MoS_2$ Driven by Acoustic Waves. *Advanced Optical Materials* **9**, 2001349 (2020).

[23] R. Peng, J. Zhu, X. Xu, and M. Li, "Control interlayer excitons in 2D heterostructures with acoustic waves," in *Conference on Lasers and Electro-Optics (CLEO)*, IEEE Conference Proceedings (2021, May): 1–2.
20

[24] Dumur, É., Satzinger, K. J., Peairs, G. A. et al. Quantum communication with itinerant surface acoustic wave phonons. *npj Quantum Information* **7**, 173 (2021).

[25] Karni, O., Barré, E., Lau, S. C. et al. Infrared interlayer exciton emission in $MoS_2/WSe_2$ heterostructures. *Physical Review Letters* **123**, 247402 (2019).

[26] Lin, Q., Fang, H., Kalaboukhov, A. et al. Moiré-engineered light-matter interactions in $MoS_2/WSe_2$ heterobilayers at room temperature. *Nature Communications* **15**, 8762 (2024).

[27] Chen, D., Dini, K., Rasmita, A. et al. Spatial filtering of interlayer exciton ground state in $WSe_2/MoS_2$ heterobilayer. *Nano Letters* **24**, 8795–8800 (2024).

[28] Streetman, B. G. and Banerjee, S. *Solid State Electronic Devices* (Prentice Hall, New Jersey, 2000).

[29] Huang, Z., Bai, Y., Zhao, Y. et al. Observation of phonon Stark effect. *Nature Communications* **15**, 4586 (2024).

[30] Barbone, M., Montblanch, A. R. P., Kara, D. M. et al. Charge-tuneable biexciton complexes in monolayer $WSe_2$. *Nature Communications* **9**, 3721 (2018).

[31] Perea-Causin, R., Brem, S., Buchner, F. et al. Electrically tunable layer-hybridized trions in doped $WSe_2$ bilayers. *Nature Communications* **15**, 6713 (2024).

[32] Li, Z., Wang, T., Lu, Z. et al. Revealing the biexciton and trion-exciton complexes in BN encapsulated $WSe_2$. *Nature Communications* **9**, 3719 (2018).

[33] Li, Z., Wang, T., Lu, Z. et al. Direct Observation of Gate-Tunable Dark Trions in Monolayer $WSe_2$. *Nano Letters* **19**, 6886-6893 (2019).

[34] Preciado, E., Schülein, F. J., Nguyen, A. E. et al. Scalable fabrication of a hybrid field-effect and acousto-electric device by direct growth of monolayer $MoS_2/LiNbO_3$. *Nature Communications* **6**, 8593 (2015).

[35] Wu, B., Zheng, H., Li, S. et al. Enhanced homogeneity of moiré superlattices in double-bilayer $WSe_2$ homostructure. *ACS Applied Materials & Interfaces* **15**, 48475–48484 (2023).

[36] Wang, Z., Chiu, Y.-H., Honz, K., Mak, K. F. and Shan, J. Electrical tuning of interlayer exciton gases in $WSe_2$ bilayers. *Nano Letters* **18**, 137–143 (2018).

[37] Huang, Z., Zhao, Y., Bo, T. et al. Spatially indirect intervalley excitons in bilayer $WSe_2$. *Physical Review B* **105**, L041409 (2022).

[38] Calman, É. V., Fowler-Gerace, L. H., Choksy, D. J. et al. Indirect Excitons and Trions in

MoSe$_2$/WSe$_2$ van der Waals Heterostructures. *Nano Letters* **20,** 1869-1875 (2020).

[39] Tan, Q., Rasmita, A., Li, S. et al. Layer-engineered interlayer excitons. *Science Advances* **7**, eabh0863 (2021).

[40] Frenkel, J. On Pre-Breakdown Phenomena in Insulators and Electronic Semi-Conductors. *Physical Review* **54**, 647-648 (1938).

[41] Nysten, E. D., Weiß, M., Mayer, B., Petzak, T. M., Wurstbauer, U., and Krenner, H. J. Scanning acousto-optoelectric spectroscopy on a transition metal dichalcogenide monolayer. *Advanced Materials* **36**, 2402799 (2024).

[42] Tongay, S., Zhou, J., Ataca, C. et al. Thermally Driven Crossover from Indirect toward Direct Bandgap in 2D Semiconductors: MoSe$_2$ versus MoS$_2$. *Nano Letters* **12**, 5576-5580 (2012).

[43] Glazov, M. M. Quantum interference effect on exciton transport in monolayer semiconductors. *Physical Review Letters* **124**, 166802 (2020).

[44] Li, C. C., Gong, M., Chen, X. D. et al. Temperature dependent energy gap shifts of single color center in diamond based on modified Varshni equation. *Diamond and Related Materials* **74**, 119-124 (2017).

[45] Wu, B., Wang, Y., Zhong, J. et al. Observation of double indirect interlayer exciton in MoSe$_2$/WSe$_2$ heterostructure. *Nano Research* **15**, 2661–2666 (2022).

[46] Molina-Sánchez, A., Palummo, M., Marini, A. and Wirtz, L. Temperature-dependent excitonic effects in the optical properties of single-layer MoS$_2$. *Physical Review B* **93**, 155435 (2016).

[47] Zhu, X., Littlewood, P. B., Hybertsen, M. S. and Rice, T. M. Exciton condensate in semiconductor quantum well structures. *Physical Review Letters* **74**, 1633 (1995).

[48] Kato, T. and Kaneko, T. Transport Dynamics of Neutral Excitons and Trions in Monolayer WS$_2$. *ACS Nano* **10**, 9687-9694 (2016).

[49] Kamban, H. C. and Pedersen, T. G. Interlayer excitons in van der Waals heterostructures: Binding energy, Stark shift, and field-induced dissociation. *Scientific Reports* **10**, 5537 (2020).

[50] Rivera, P., Seyler, K. L., Yu, H. et al. Valley-polarized exciton dynamics in a 2D semiconductor heterostructure. *Science* **351**, 688–691 (2016).

[51] Liu, Y., Elbanna, A., Gao, W., Pan, J., Shen, Z. and Teng, J. Interlayer excitons in

transition metal dichalcogenide semiconductors for 2D optoelectronics. *Advanced Materials* **34**, 2107138 (2022).

[52] Massicotte, M., Vialla, F., Schmidt, P. et al. Dissociation of two-dimensional excitons in monolayer $WSe_2$. *Nature Communications* **9**, 1633 (2018).

[53] Smidstrup, S., Markussen, T., Vancraeyveld, P. et al. QuantumATK: an integrated platform of electronic and atomic-scale modelling tools. *Journal of Physics-Condensed Matter* **32**, 36 (2020).

[54] Perdew, J. P., Burke, K. and Ernzerhof, M. Generalized gradient approximation made simple. *Physical Review Letters* **78**, 1396-1396 (1997).

[55] Schlipf, M. and Gygi, F. Optimization algorithm for the generation of ONCV pseudopotentials. *Computer Physics Communications* **196**, 36-44 (2015).

[56] Ferreira, L. G., Marques, M. and Teles, L. K. Approximation to density functional theory for the calculation of band gaps of semiconductors. *Physical Review B* **78**, 125116 (2008).

[57] Guilhon, I., Koda, D. S., Ferreira, L. G., Marques, M. and Teles, L. K. Approximate quasiparticle correction for calculations of the energy gap in two-dimensional materials. *Physical Review B* **97**, 045426 (2018).

[58] HExd, J., Scuseria, G. E. and Ernzerhof, M. Hybrid functionals based on a screened Coulomb potential. *Journal of Chemical Physics* **118**, 8207-8215 (2003).

[59] Krukau, A. V., Vydrov, O. A., Izmaylov, A. F. and Scuseria, G. E. Influence of the exchange screening parameter on the performance of screened hybrid functionals. *Journal of Chemical Physics* **125**, 224106 (2006).

[60] Grimme, S. Semiempirical GGA-type density functional constructed with a long-range dispersion correction. *Journal of Computational Chemistry* **27**, 1787-1799 (2006).